Limitations of empirical sediment transport formulas for shallow water and their consequences for swash zone modelling

WEI LI, Lecturer, *Ocean College, Zhejiang University, Hangzhou, People's Republic of China*

*Email: lw05@zju.edu.cn*

PENG HU, Associate Professor, *Ocean College, Zhejiang University, Hangzhou, People's Republic of China*

*Email: pengphu@zju.edu.cn (author for correspondence)*

THOMAS, PAHTZ, Professor, Ocean College, Zhejiang University, Hangzhou, People's Republic of China

*Email: tpaehtz@gmail.com*

ZHIGUO HE, Professor, Ocean College, Zhejiang University, Hangzhou, People's Republic of China

*Email: hezhiguo@zju.edu.cn*

ZHIXIAN CAO, Professor, State Key Laboratory of Water Resources and Hydropower Engineering Science, *Wuhan University, People's Republic of China*

*Email: zxcao@whu.edu.cn*

# Limitations of empirical sediment transport formulas for shallow water and their consequences for swash zone modelling

**ABSTRACT**

Volumetric sediment concentrations computed by phase-resolving swash morphodynamic models are shown to exceed unity minus porosity (i.e., the maximally physically possible concentration value) by up to factor $10^5$ when using standard expressions to compute the sediment transport rate. An ad hoc limit of sediment concentration is introduced as a means to evaluate consequences of exceeding physically realistic concentration by standard expressions. We find that implementation of this ad hoc limit strongly changes the quantitative and qualitative predictions of phase-resolving swash morphodynamic models, suggesting that existing swash predictions are unreliable. This is because standard expressions inappropriately consider or ignore the fact that the shallow swash water depth limits the storage capacity of transported sediment.

## 1  Introduction

The swash zone is characterized by very shallow water and successively covered and uncovered by water due to wave run-up/backwash. In numerical and analytical studies of swash sediment transport, standard expressions are used to compute the sediment transport rate (e.g. Briganti, Dodd, Pokrajac, & O'Donoghue, 2012; Kelly, & Dodd, 2010; Postacchini, Brocchini, Mancinelli, & Landon 2012; Postacchini, Othman, Brocchini, & Baldock, 2014; Zhu, Dodd, & Briganti, 2012; Zhu, & Dodd, 2013a, 2013b, 2015), which, however, are not derived or calibrated for very shallow swash conditions. In fact, most, if not all, standard expressions for the sediment transport rate are either independent of or inversely proportional to the water depth (Zhu, & Dodd, 2013a). For very small water depths, this may lead to sediment transport rates so large that the corresponding sediment concentration may be larger than it can physically be. As an attempt to overcome this potential issue, Pritchard, & Hogg (2005) multiplied the standard Grass expression by the water depth (termed as PH expression below). However, whether the PH expression resolves the issue with small water depth remains unexamined. Moreover, some studies avoided this issue by phase-averaging the uprush and backwash of the swash (e.g. Masselink, & Hughes, 1998; Hughes, Masselink, & Brander, 1997). However, implementation of phase-averaging prevents models from predicting detailed morphological evolution processes.

Here existing phase-resolving analytical and numerical studies of swash morphydynamics are revisited. In particular, we examine the magnitudes of sediment concentration, which appears implicitly in standard sediment transport formulas. We then impose an ad-hoc upper limit on the sediment concentration and thereby limit the sediment transport rate. It is strongly emphasized here that the introduction of this ad-hoc limit is **NOT** meant to lead to a better "model". Instead it is only a means to test whether the exceeding physically realistic sediment concentrations has serious consequences on the overall predictions of phase-resolving swash morphydynamic models.

## 2    Analytical investigations

We revisit the theoretical analysis of swash beach evolution by Kelly, & Dodd (2010). They used the swash solution derived by Shen, & Meyer (1963) to represent the swash hydrodynamics, which reads:

$$h(x,t) = \frac{[4\sqrt{gh_0}t - gt^2\tan\alpha - 2x]^2}{36gt^2} \qquad (1a)$$

$$u(x,t) = \frac{2[\sqrt{gh_0}t - gt^2\tan\alpha + x]}{3t} \qquad (1b)$$

where $h$ is the water depth; $u$ is the swash velocity; $x$ is the horizontal distance (positive landwards), $t$ is time; $g$ is gravitational acceleration, $\tan\alpha$ is the beach slope, $h_0$ is the initial wave height. The Exner equation that governs beach deformation reads:

$$(1-p)\frac{\partial z}{\partial t} + \frac{\partial q_b}{\partial x} = 0 \qquad (2)$$

where $q_b$ is the bed load transport rate, $z$ is the beach elevation, $p$ is the sediment porosity. Beach deformation depth over a swash cycle at a given position $x_p$ is obtained by integrating Eq. (2), which, using the standard Grass expression $q_b = A_G u^3$ and Eq. (1b) for swash velocity, results in:

$$\Delta z(x_p) = \int_{t_i}^{t_d} \partial z = \frac{1}{p-1}\int_{t_i}^{t_d}\frac{\partial q_b}{\partial x}dt = \frac{3A_G}{p-1}\int_{t_i}^{t_d}(u^2\frac{\partial u}{\partial x})dt = \frac{8A}{9(p-1)}\int_{t_i}^{t_d}\frac{[\sqrt{gh_0}t - gt^2\tan\alpha + x]^2}{t^3}dt \quad (3)$$

where $t_i$ and $t_d$ are the inundation and denudation times, which can be derived by setting $h = 0$ and $x = x_p$ in Eq. (1a). The resulted final expression reads:

$$\Delta z(x_p) = \frac{1}{p-1}\frac{8A_G}{9}[3g\sqrt{4h_0^2 - 2h_0 x_p \tan\alpha} + (gh_0 - 2x_p g \tan\alpha)\ln(t_d/t_i)] \quad (4)$$

For comparison purposes, a modified expression for sediment transport rate is introduced by imposing an upper limit on the sediment concentration, which reads:

$$q_b = \min(huc_{upper}, A_G u^3) \quad (5)$$

where $c_{upper}$ is the maximally physically possible concentration value defined as $c_{upper} = 1-p$. This definition means that the entire flow is effectively filled with sediment. We wish to strongly emphasize that this technical note does NOT propose Eq. (5) as a better 'model'. Instead, it is only introduced to determine which predictions of the standard Grass expression (and further standard expressions that we test later) are due to exceeding the maximally physically possible sediment concentration. In fact, any difference between model predictions when using the standard Grass expression on the one hand and the modified Grass expression (Eq. (5)) on the other hand MUST be a result of exceeding this maximum and thus has no physical justification. It is worthy pointing out that limiting the maximal sediment concentration has been done before: Garcia, & Parker (1991) derived an empirical relation for suspended sediment concentration with an upper limit of 0.3 [approximately half of $(1-p)$].

When the modified expression Eq. (5) is used, the beach deformation is obtained by numerically integrating Eq. (2), which gives:

$$\Delta z(x_p) = \frac{1}{p-1}\delta t \sum_{j=1}^{n}\frac{q_b(x_p + \delta x, t_i + j\delta t) - q_b(x_p, t_i + j\delta t)}{\delta x} \quad (6)$$

In Eq. (6), the period between $t_i$ and $t_d$ is divided into $n$ intervals of length $\delta t$; $\delta x$ is a sufficiently small distance.

We use the same parameter values as in Kelly, & Dodd (2010): $p = 0.4$; $A_G = 0.004$ s²/m; $h_0 = 0.65$ m; $\tan\alpha = 0.1$. The period $T \equiv 4\sqrt{gh_0}/(g\tan\alpha) = 10.3$ s. Figure 1 illustrates the final beach deformation depth (Fig. 1a) and beach profiles at three instants (Fig. 1b, Fig. 1c and Fig. 1d). In Fig. 1, Run 1 represents the analytical solution by Eq. (4), Run 2 is the numerical integration of Eq. (6) fed by Eq. (3), and Run 3 (limited sediment transport rate) is the numerical integration of Eq. (6) fed by Eq. (5). From Fig. 1a, Run 1 and Run 2 agree satisfactorily with each other, indicating the high accuracy of the numerical integration. Thus Run 1 and Run 2 are combined later in Fig. 1b, Fig. 1c and Fig. 1d. Big differences are seen between Run 1/2 and Run 3. Run 1/2 computes bed degradation everywhere in the swash zone, consistent with Kelly, & Dodd (2010). A so-called "bed step" is computed at the swash front (Run 1/2, Fig. 1b, Fig. 1c and Fig. 1d), but it is washed out at the end of the swash (Run 1/2, Fig. 1a). In contrast, the bed step is absent for Run 3. Transient bed aggradations can be seen in Run 3 in some regions (Fig. 1b, Fig. 1c and Fig. 1d).

These differences between Run 1/2 and Run 3 are explained using Fig. 2, which shows the cross-shore variation of the swash depth (Fig. 2a) and velocity (Fig. 2c) and concentrations (Fig. 2b) computed with the standard Grass expression. From Fig. 2b, the computed concentrations at the end of the uprush (at about 5.15 s) are within the physically reasonable range. This is because, at the reverse of the swash, the swash is characterized by the smallest velocities (Fig. 2c). However, concentrations at both the uprush and backwash exceed the maximal physically possible concentration value (2 s represents uprush and 8 s represents backwash). As the swash depth decreases up to the beach, the concentration increases rapidly and attains values as high as $5 \times 10^4$, which is more than $10^5$ times larger than the physically reasonable range. This is because the standard Grass expression does not account for the effects of the very shallow swash water depth.

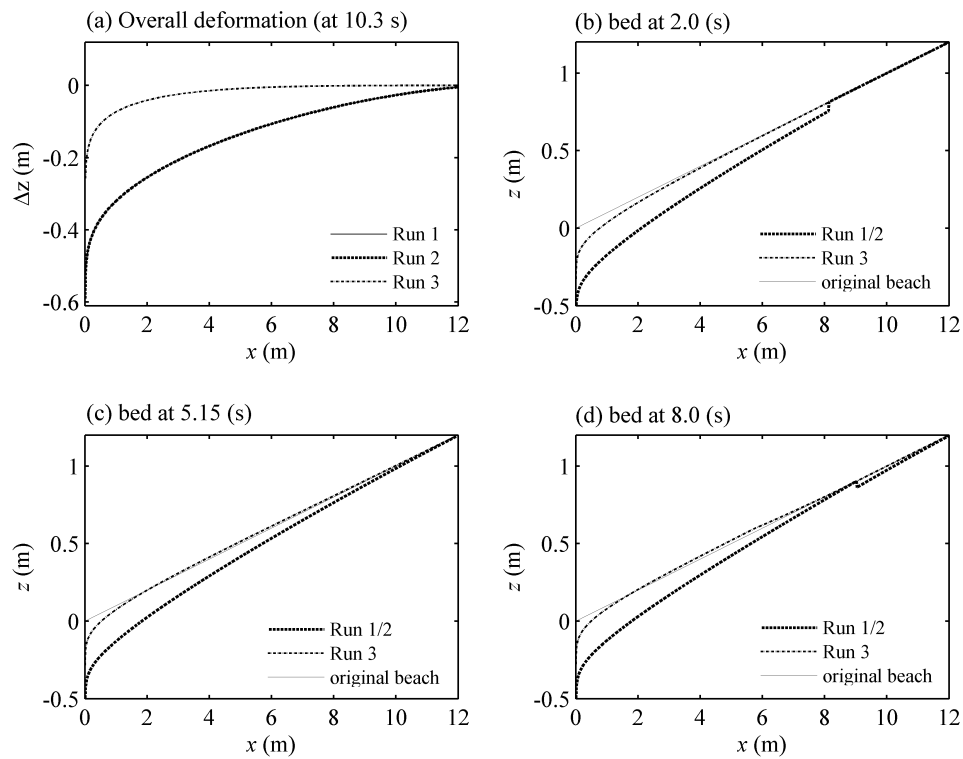

Figure 1. Cross shore variations of (a) beach deformation depth, and (b, c, d) beach profiles. Run 1 and 2 correspond to using the standard Grass expression to compute the sediment transport rate and Run 3 to using the Grass expression modified by the upper concentration limit. Also included in Fig. 1b, Fig. 1c and Fig. 1d is the original beach profile.

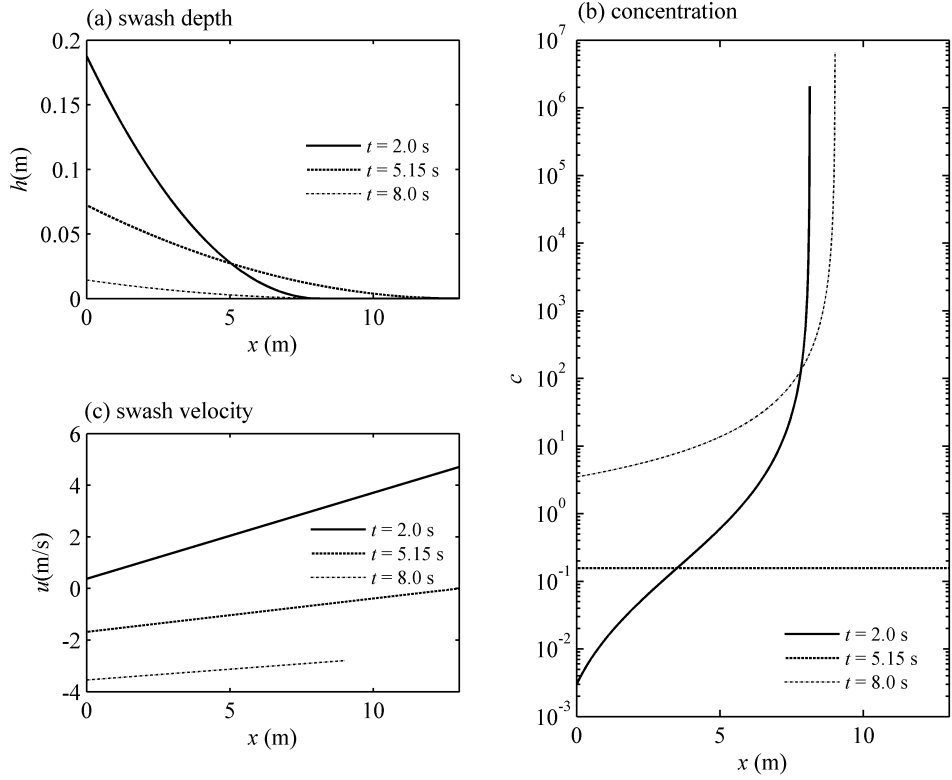

Figure 2. Cross-shore variations of (a, c) the swash hydrodynamics, and (b) the volumetric sediment concentration. Sediment transport rates are computed using the standard Grass expression.

## 3 Numerical Investigations

The coupled phase-resolving swash morphological model of Hu, Li, He, Pahtz, & Yue (2015) is applied. The governing equations read

$$\frac{\partial h}{\partial t} + \frac{\partial hu}{\partial x} = 0 \tag{7}$$

$$\frac{\partial hu}{\partial t} + \frac{\partial (hu^2 + 0.5gh^2)}{\partial x} = -gh\frac{\partial z}{\partial x} - c_D u|u| \tag{8}$$

$$\frac{\partial (1-p)z}{\partial t} + \frac{\partial q_b}{\partial x} = 0 \tag{9}$$

where $c_D$ is the drag coefficient. Eqs. (7, 8. 9) are solved by a shock-capturing finite volume method along with a well-balanced version of the Slope Limited Centered scheme (Hu et al., 2015). The numerical cases are designed following Zhu, & Dodd (2013a). On a beach of uniform slope $\tan\alpha = 0.1$, beach elevation $z(x) = x\tan\alpha$. The region with $x \geq 0$ is initially dry. On the seaside ($x<0$): $h_0 = 0.65$ m. The swash is simulated by allowing the water on the seaside to freely flow up and down the beach. The spatial step $\Delta x = 0.005$ m.

Figure 3 presents the non-dimensional beach deformation depth ($\Delta z/(h_0 \tan^2 \alpha)$) in the space-time plane computed with the standard (Fig. 3a) and modified (Fig. 3b) Grass expressions. When the standard Grass expression is applied (Fig. 3a), the beach degrades in the lower swash region and aggrades in the upper swash region during run-up, whereas it degrades in the whole swash zone during backwash. A bed step is computed during the uprush, but it is washed out at the end of the swash. These results are consistent with Zhu, & Dodd (2013a). While bed steps do occur in nature, the explanation of their formulation obviously cannot rely on expressions that compute sediment transport rates much larger than maximally possible values. When the modified Grass expression is implemented (Fig. 3b), the following points are noted. First, the magnitudes of both deposition and erosion are reduced, which follows naturally from limiting the sediment transport rate. Second, the area with deposition is wider, and the final bed profile is characterized by some deposition. This indicates some extent of onshore sediment transport, which is consistent with field studies (Masselink, Evans, Hughes, & Russell, 2005; Kelly, & Dodd, 2010). In contrast, using the standard Grass expression always leads to net offshore transport everywhere on the beach (bed degradation everywhere after a swash cycle), which is at odds with field studies (Kelly, & Dodd, 2010). Third, a bed step is absent during the uprush. Instead, a bed hump is left on the beach at the end of the swash (see the contours of -0.1, 0.005, 0.08 in Fig. 3b). These results indicate strong limiting effects of the small swash water depth on beach morphological changes.

To evaluate the influence of the small swash depth more comprehensively, extensive numerical cases studies are conducted, covering a wide range of bed frictions, bed mobility and sediment empirical expressions (the modifications of these formulas with an upper concentration limit are done analogous to the modification of the Grass expression). Comparisons are made between the cases with and without modifications through a relative overall discrepancy $R$ defined as:

$$R(\Delta z) = \frac{\sum_{i,j}|\Delta z(x_i,t_j) - \Delta \overset{1}{z}(x_i,t_j)|}{\sum_{i,j}|\Delta z(x_i,t_j)|} \qquad (10)$$

where $\Delta z(x_i,t_j)$ and $\Delta \overset{1}{z}(x_i,t_j)$ are the computed beach deformation depth at position $x_i$ and time $t_j$ using the standard and modified expressions, respectively. The higher the R value, the larger the effects of exceeding the upper limit of concentration are.

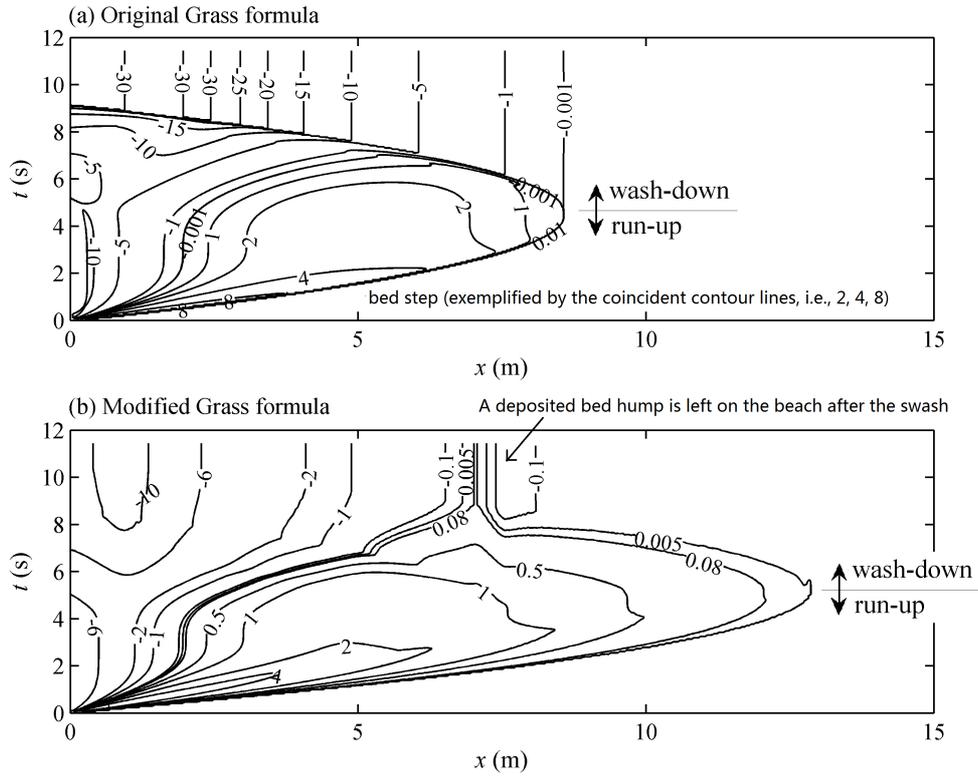

Figure 3. Computed non-dimensional beach deformation depth in the space-time plane with (a) $q_b = A_G u^3$, and (b) $q_b = \min[hu(1-p), A_G u^3]$. $A_G$ =0.004 s2/m and $c_D = 0.0$.

Figure 4 shows the variation of the R-value with the bed mobility in relation to (a) the Grass expression, and (b, c, d) the PH expression ($q_b = A_{PH} h^k u^3$) with the exponent $k$ = 0.8 (Fig. 4b), 1.0 (Fig. 4c) and 1.3 (Fig. 4d). The bed mobility $\sigma = A_G g(1-p)$ for the Grass expression, and $\sigma = A_{PH} g(1-p) h^k$ for the PH expression. The comparisons in relation to the Bagnold, MPM, Van Rijn, and Bailard expressions (Zhu, & Dodd, 2013a) are similar to Fig. 4a and thus are not shown. From Fig. 4a, the R-value is consistently around 50% for a frictionless beach ($c_D = 0.0$). This means the standard expressions produce significant errors for a frictionless beach, which unfortunately has been a common assumption in most theoretical swash studies. When the beach friction is considered, these standard expressions still produce significant errors for a wide range of bed mobility and friction conditions. A decreasing trend of R with lower bed mobility is clear from Fig. 4a. This suggests that reducing the coefficient (i.e. $A_G$ for the standard Grass expression) through calibration may ensure physically realistic sediment transport rates. However, in some cases, the coefficient would have to be reduced by a magnitude of $10^5$. This would imply negligible or even no sediment motion and thus be of no practical interest. Kelly, & Dodd (2010) has implied $0.01 < \sigma < 0.2$ for sand beaches, which prohibits a strong reduction of the coefficient. When the PH expression is used, the extent of exceeding physically reasonable sediment concentrations is largely reduced, as reflected by the smaller R-value (Fig. 4b, Fig. 4c and Fig. 4d). This is due to the positive correlation between water depth and sediment transport rate in the PH expression. However, the R-value is not vanishing at all. Note that a larger $k$ in the PH expression would lead to a smaller R-value, which means that a sufficiently large $k$ could give physically reasonable concentration. However, this kind of mathematical play is physically meaningless.

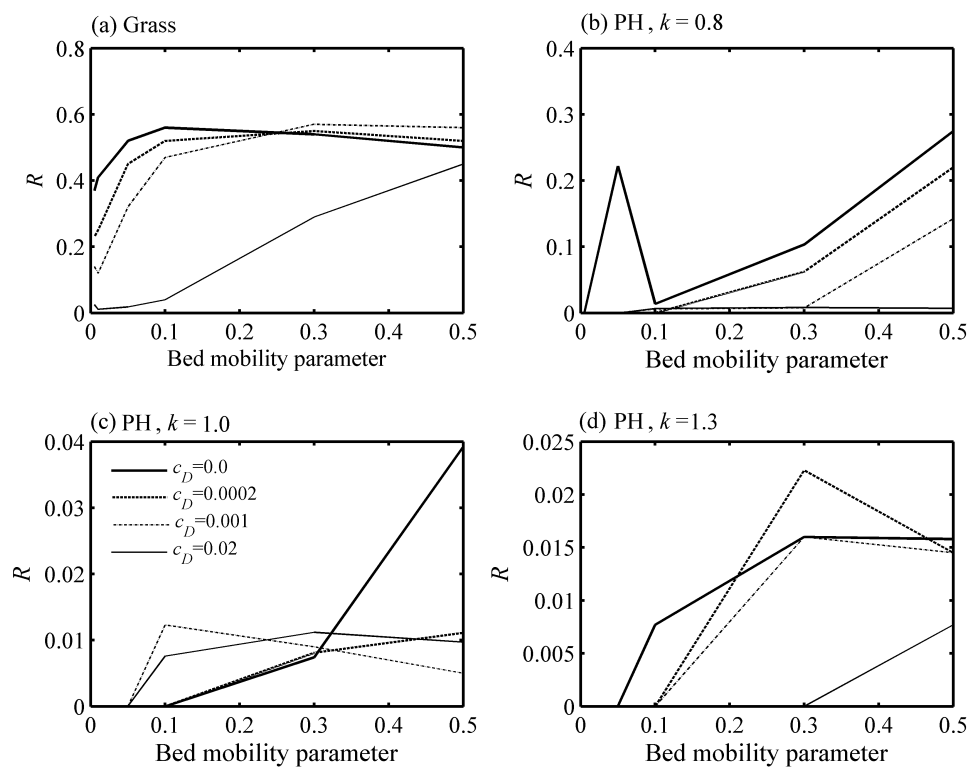

Figure 4. Relative discrepancy of the computed beach deformation depth in relation to (a) the Grass expression, (b, c, and d) the PH expression with different exponents.

## 4 Conclusions

Beach deformation in the swash zone is studied using phase-resolving analytical (Section 2) and numerical (Section 3) approaches. Two distinct cases were compared: the standard cases that consider standard expressions for the sediment transport rate, and the modified cases in which an upper limit of the sediment concentration is imposed. The modified cases yielded results quantitatively and even qualitatively different from the standard cases. In particular, our study indicates that the bed step computed by many morphodynamic swash zone models might be an artifact of exceeding physically realistic sediment transport rates. Indeed, the main reason why the maximal physically possible values of the sediment transport rate are exceeded so often (by up to factor $10^5$ in some cases) is the small swash depth, of which the effect is not accounted for in standard expressions for the sediment transport rate.

Though previous studies have noted many issues with the standard expressions (Bagnold, 1966; Pritchard, & Hogg, 2005), only a few of them have ever considered the issue of the vanishing water depth and its effect on sediment transport rates. One exception is Pritchard, & Hogg (2005), in which the Grass expression was multiplied by the swash depth. Yet, a detailed examination of the PH expression shows that it does not account for vanishing water depth in a satisfactory manner. Finally, we would like to note that, when other swash analytical solutions, which for example are affected by seaward boundary conditions (Guard, & Baldock, 2007), were implemented, the present findings should hold true because the very small swash depth is a common feature (Shen, & Meyer, 1963).

**Acknowledgements**

The authors are very grateful to several anonymous reviewers for their constructive and critical comments that greatly improved the paper.

**Funding**

This research was partly supported by the Research Fund for Doctoral Program of Higher Education of China (grant 20130101120152) and the National Natural Science Foundation of China (grants 11402231, 11502233 and 41376095).

**Notation**

$A_G$ $A_{PH}$ = empirical coefficients in sediment transport relations (-)

$c_D$ = drag coefficient (-)

$c_{upper}$ = maximally physically possible concentration value (-)

$g$ = gravitational acceleration (m s$^{-2}$)

$h$ = water depth (m)

$h_0$ = initial wave height (m)

$k$ = empirical exponent in the PH expression (-)

$p$ = sediment porosity (-)

$q_b$ = unit-width volumetric bed load transport rate (m² s⁻¹)

$R$ = relative overall discrepancy (-)

$t$ = time (s)

$t_i$, $t_d$ = inundation and denudation times (s)

$T$ = period of a swash cycle (s)

$u$ = swash velocity (m s⁻¹)

$x$ = horizontal distance (m)

$z$ = beach elevation (m)

$\tan\alpha$ = beach slope (-)

$\sigma$ = bed mobility number

$\Delta x$ = spatial step (m)

$\Delta z$ = beach deformation depth (m)

**References**

Bagnold, R. A. (1966). *An approach to the sediment transport problem from general physics*. Professional Paper 422-1, Washington, D.C. Retrieved from USGS website: http://pubs.usgs.gov/pp/0422i/report.pdf

Briganti, R., Dodd, N., Pokrajac, D., & O'Donoghue, T. (2012, July). *Numerical and experimental description of the flow, boundary layer and bed evolution in bore-driven swash on a coarse sediment beach*. Paper presented at Proceedings of the 33rd Conference on Coastal Engineering, Santander, Spain.

Garcia, M., & Parker, G. (1991). Entrainment of bed sediment into suspension. *Journal of Hydraulic Engineering ASCE, 117(4)*, 414-435.

Guard, P. A. & Baldock, T. E. (2007). The influence of seaward boundary conditions on swash zone hydrodynamics. *Coastal Engineering, 54(4)*, 321-331.

Hu, P., Li, W., He, Z., Pähtz, T., & Yue, Z. (2015). Well-balanced and flexible modelling of swash hydrodynamics and sediment transport. *Coastal Engineering, 96(2),* 27-37.

Hughes, M. G., Masselink, G., & Brander, R. W. (1997). Flow velocity and sediment transport in the swash one of a steep beach. *Marine Geology, 138(1-2)*, 91-103.

Kelly, D. M., & Dodd, N. (2010). Beach face evolution in the swash zone. *Journal of Fluid Mechanics, 661,* 316-440.

Masselink, G., & Hughes, M. (1998). Field investigation of sediment transport in the swash zone. *Continental Shelf Research, 18(10),* 1179-1199.

Masselink, G., Evans, D., Hughes, M. G., & Russell, P. (2005). Suspended sediment transport in the swash zone of a dissipative beach. *Marine Geology, 216(3),* 169-189.

Postacchini, M., Brocchini, M., Mancinelli, A., & Landon, M. (2012). A multi-purpose, intra-wave, shallow water hydro-morphodynamic solver. *Advances in Water Resources, 38(3),* 13-26.

Postacchini, M., Othman, I. K., Brocchini, M., & Baldock, T. E. (2014). Sediment transport and morphodynamics generated by a dam-break swash uprush: coupled vs uncoupled modeling. *Coastal Engineering, 89(7),* 99-105.

Pritchard, D., & Hogg, A. J. (2005). On the transport of suspended sediment by a swash event on a plane beach. *Coastal Engineering, 52(1),* 1-23.

Shen, M. C., & Meyer, R. E. (1963). Climb of a bore on a beach, III: run-up. *Journal of Fluid Mechanics, 16,* 113-125.

Zhu, F. F., Dodd, N., & Briganti, R. (2012). Impact of a uniform bore on an erodible beach. *Coastal Engineering, 60(2),* 326-333.

Zhu, F. F., & Dodd, N. (2013a). Net beach change in the swash zone: a numerical investigation. *Advances in Water Resources, 53(3),* 12-22.

Zhu, F. F., & Dodd, N. (2013b, June). *The influence of bed friction on the creation of a bed-step by a solitary wave in the swash*. Paper presented at the 7th International Conference on Coastal Dynamics, Arcachon Convention Centre, France.

Zhu, F. F., & Dodd, N. (2015). The morphodynamics of a swash event on an erodible beach. *Journal of Fluid Mechanics, 762*, 110-140.